\documentclass[aps,prl,reprint,superscriptaddress]{revtex4-2}
\usepackage{amsmath}
\usepackage{amssymb}
\usepackage{longtable}
\usepackage{graphicx}
\usepackage{multirow}
\usepackage{makecell}
\usepackage{hyperref}
\usepackage{float}
\usepackage{url}
\usepackage{epstopdf}
\usepackage{soul,color,xcolor}

\begin{document}
\title{Autferroics-based true random number generators with enhanced performance}
\author{Jun-Jie Zhang}
\affiliation{Key Laboratory of Quantum Materials and Devices of Ministry of Education, School of Physics, Southeast University, Nanjing 211189, China}
\affiliation{Department of Materials Science and NanoEngineering, Rice University, Houston, Texas 77005, USA}
\author{Shuai Dong}
\email{sdong@seu.edu.cn}
\affiliation{Key Laboratory of Quantum Materials and Devices of Ministry of Education, School of Physics, Southeast University, Nanjing 211189, China}
\author{Boris I. Yakobson}
\email{biy@rice.edu}
\affiliation{Department of Materials Science and NanoEngineering, Rice University, Houston, Texas 77005, USA}

\begin{abstract}
Physical entropy-driven true random number generators are essential for emerging probabilistic computing paradigms, but conventional implementations based on magnetic tunneling junctions have reached their performance plateaus limited by inherent tradeoffs and weak tunability. Here, autferroics, a sister branch of multiferroics, is proposed to construct true random number generators. Benefiting from its unique energy landscape due to strong seesaw-type magnetoelectricity, the performance of true random number generation can be significantly enhanced in autferroic tunneling junctions, verified by passing standard statistical tests. Furthermore, autferroics-based random number generators can exhibit multi-field tunability and intrinsic multi-state randomness, opening an avenue for efficient hardware realization of the complex-number arithmetic and simulation of probability distributions of quantum mixed states within stochastic circuits.
\end{abstract}
\maketitle

True random number generators (TRNGs) that derive entropy directly from physical fluctuations are essential for high-security cryptography, stochastic computing, and neuromorphic hardware  \cite{1,2,3,4,5,6,7}. Unlike pseudo-random number generation algorithms, physical TRNGs are inherently unpredictable \cite{7}, but most existing implementations are limited by their low throughput and high power consumption \cite{6,vodenicarevic2017low}. Recently, thermally unstable magnetic tunnel junctions (MTJs) have emerged as attractive candidates for both high-speed TRNGs and probabilistic computers ($p$-computers) \cite{9,10,11,12,13,14,15,kaiser2021probabilistic}. The performance of MTJ-based probabilistic TRNGs is influenced by several key factors, including tunnel magnetoresistance, fluctuating switching rates, and tunable randomness \cite{11,16}. However, these factors may compete with each other\cite{9,10}, making them difficult to balance within conventional single-order magnetic systems \cite{16}. Hence, alternative physical mechanisms are crucial to pursue high-performance TRNGs and $p$-computers for advancing unconventional computing.

Recently, autferroicity has been proposed as a sister form of multiferroicity, which exhibits intrinsically strong seesaw-type magnetoelectric (ME) coupling \cite{17,18,19}. Although autferroicity has not been experimentally confirmed yet, similar strong competition between ferroelectric and magnetic orders was experimentally observed in $[1-x]$(Ca$_{0.6}$Sr$_{0.4}$)$_{1.15}$Tb$_{1.85}$Fe$_{2}$O$_{7}-[x]$Ca$_{3}$Ti$_{2}$O$_{7}$ series \cite{pitcher2015tilt} and also theoretically predicted in hole-doped CuInP$_2$S$_6$ \cite{zhang2023origin}, without revealing autferroicity concept. Unlike multiferroics \cite{19,20,21}, autferroics do not require the simultaneous coexistence of magnetic and ferroelectric orders in a phase. Instead, their magnetic or ferroelectric order appears individually in a phase and the strong mutual exclusion between these two orders can be utilized in functional applications. For example, this inherent exclusion, arising from the strong ME coupling, can substantially reduce its effective magnetic (ferroelectric) switching barrier, thereby promoting thermally activated fluctuations. More importantly, this seesaw-type ME behavior decouples these two order parameters, enabling independent control over their individual switching dynamics, as well as their equilibrium state populations, which is conceptually different from the multiferroic systems.

The unitive phase diagram of autferroics and (type-I) multiferroics can be quantitatively described in terms of Landau theory \cite{18}. Independent of specific crystal symmetry, symmetry constraints imply that the lowest-order ME term is biquadratic: $cP^2M^2$ with $c>0$, where $P$ and $M$ are the ferroelectric and magnetic order parameters, respectively. In the Landau theory \cite{18,22}, there are two critical thresholds: $c_M\equiv2F_M/P_0^2M_0^2$ and $c_P\equiv2F_P/P_0^2M_0^2$, where $F_M$ ($F_P$) is the switching barrier and $M_0$ ($P_0$) is the spontaneous magnetic (polar) moments of the pure magnetic (ferroelectric) system without ME coupling. The autferroic phase appears as the ground state when ME coupling is strong enough (i.e. $c>c_M$ and $c>c_P$), where four wells (ferroelectric $vs.$ magnetic) exist in its energy landscape. Multiferroic and single-ferroic phases will emerge when coupling is relatively weak [see End Matter (EM) for details].

In this Letter, we show that autferroics can be utilized to construct TRNGs, which can substantially enhance the performance compared to current mainstream TRNGs based on MTJs. Benefiting from its ME coupling, its generation rate can be further improved by the electrical field while preserving near-ideal stochasticity. Moreover, if the ferroelectric and magnetic phases in autferroics are nearly degenerate (i.e. $c_M\approx c_P$), all four states are statistically accessible, enabling the direct realization of a quaternary TRNG. Based on these features, the complex-number arithmetic and simulation of probability distributions of quantum mixed states will be available in stochastic circuits.

\begin{figure}
\centering
\includegraphics[width=0.45\textwidth]{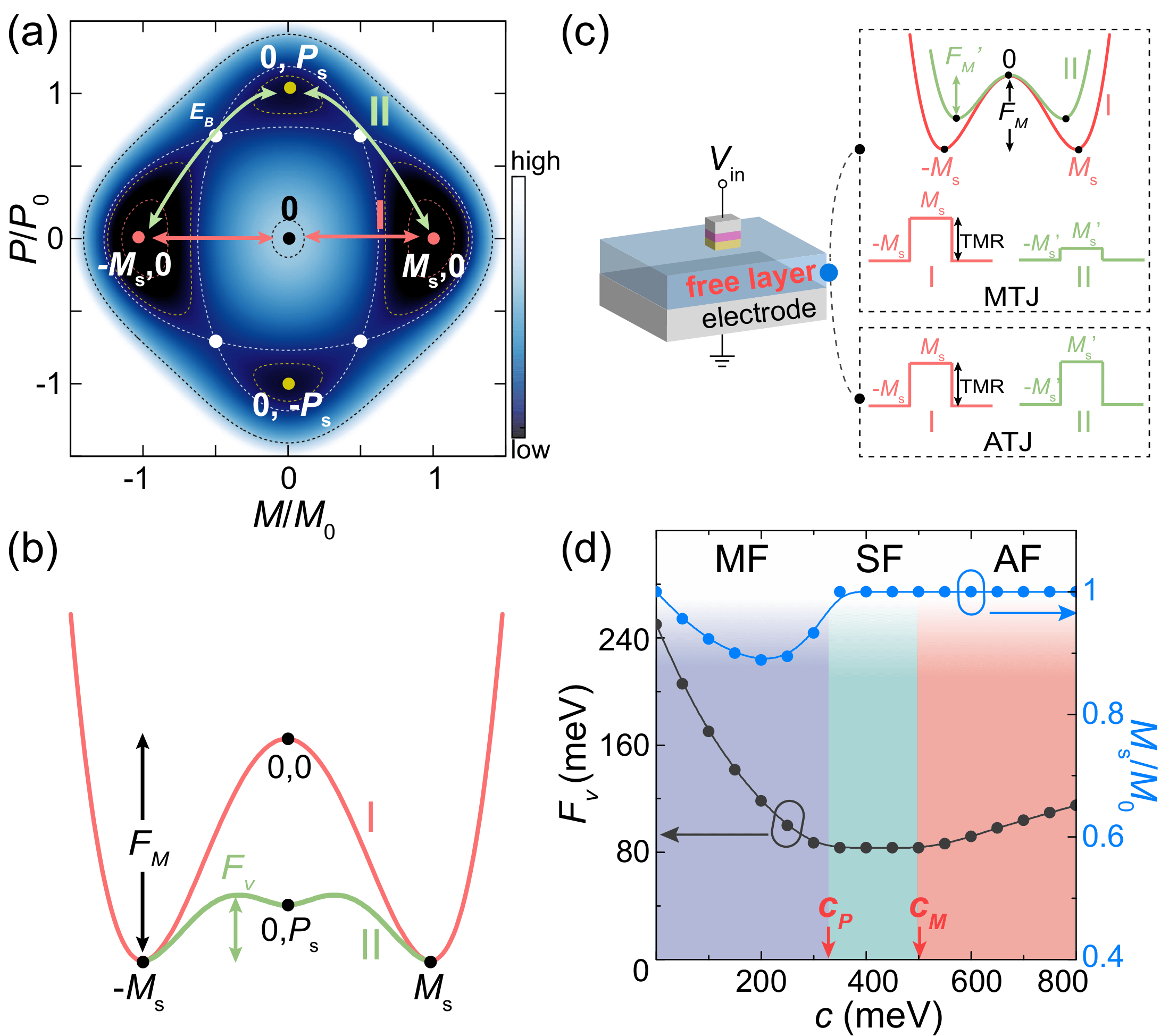}
\caption{(a) Schematic Landau free energy landscape $F(M, P)$ of autferroic system. There are four local energy minimums: ($0$, $\pm P_s$) and ($\pm M_s$, $0$). $P_s$ and $M_s$ are spontaneous polarization and magnetization, respectively. (b) The energy curves of two possible switching paths between ($+ M_s$, $0$) and ($- M_s$, $0$). The path I is ($- M_s$, $0$) $\leftrightarrow$ ($0$, $0$) $\leftrightarrow$ ($+ M_s$, $0$); the path II is ($- M_s$, $0$) $\leftrightarrow$ ($0$, $\pm P_s$) $\leftrightarrow$ ($+ M_s$, $0$). The path II with an energy barrier $F_v$, circumvents the high barrier $F_M$ of path I. These two paths are also indicated by double-headed arrows in (a). (c) Sketch of a tunnel junction device. The $free~layer$ can be pure magnetic or autferroic (i.e. MTJ $vs$ ATJ). Right: for pure magnetic layer, the reduced switching barrier (from red to green) is beneficial to the generation rate of true random numbers, but the TMR signal quality (i.e. the contrast between $\pm M_s$) will be reduced accordingly. For ATJ, the TMR signal quality will not be altered when the switching is along the energy-barrier-reduced path II. (d) Switching barrier $F_v$ and $M_s$ as a function of ME coupling $c$. MF: the type-I multiferroic; SF: single-ferroic (i.e. magnetic if $c_M>c_P$); AF: autferroic.}
\label{Fig1}
\end{figure}

With the four-well free-energy landscape, there are two different switching paths between the ($+M_s$, 0) and ($-M_s$, 0) states. As shown in Figs.~\ref{Fig1}(a-b), the effective magnetic switching barrier is $F_v$ (the path II), lower than the barrier $F_M$ of pure magnetic switching (the path I). Such a reduced effective barrier can be utilized to construct high-performance TRNGs based on autferroic tunnel junctions (ATJs), mimicking the MTJ counterparts, as compared in Fig.~\ref{Fig1}(c). In MTJ-based TRNGs \cite{22}, the generation rate of random tunnel magnetoresistance (TMR) signal would be enhanced by reducing the volume of magnetic layer or applying a transverse magnetic field \cite{11,12}. However, these approaches will also reduce the swing magnitude between $\pm M_s$, thereby decreasing the TMR variation and making signal quality worse. In contrast, the TMR signal in ATJ is preserved since the $\pm M_s$ states in free layer are not changed [Fig.~\ref{Fig1}(c)], while the effective switching barrier is substantially reduced through the ME coupling [see Supplementary Materials (SM) \cite{22}]. This superiority makes autferroics promising candidates to be implemented in hardware-level TRNGs.

In the following, we systematically investigated key parameters that may influence the performance of autferroics-based TRNGs, including the ME coupling strength $c$, the free-energy difference $\Delta F$ between the ferroelectric and magnetic phases, and the spatial gradient coefficients $\lambda_M$ and $\lambda_P$. Hence, the conclusions of this work are general and applicable to other autferroic systems. For $c_M>c_P$ case, e.g., TiGeSe$_3$ monolayer with $c_M=500$ meV, $c_P=334$ meV, and $c=550$ meV \cite{17,18}, the magnetic phase serves as the entropy source in TRNG, while the metastable ferroelectric state plays the role of intermediate (here $P_0$ and $M_0$ are normalized to $1$ which is also assumed through the rest of this study). Here, we first characterize the intrinsic energy landscape in the idealized single-unit-cell limit. It should be noted that these predicted rates can be considered as an upper-limit. The calculated effective barrier is significantly decreased from $F_M=250$ meV (obtained at $c=0$ meV, corresponding to the path I) to $F_v=87$ meV (at $c=550$ meV, the path II) [Fig.~\ref{Fig1}(d)]. As aforementioned, the ME coupling has negligible effect on the value of $\pm M_s=\pm M_0$ [Fig.~\ref{Fig1}(d)], which will preserve the quality of TMR signal.

\begin{figure}
\centering
\includegraphics[width=0.48\textwidth]{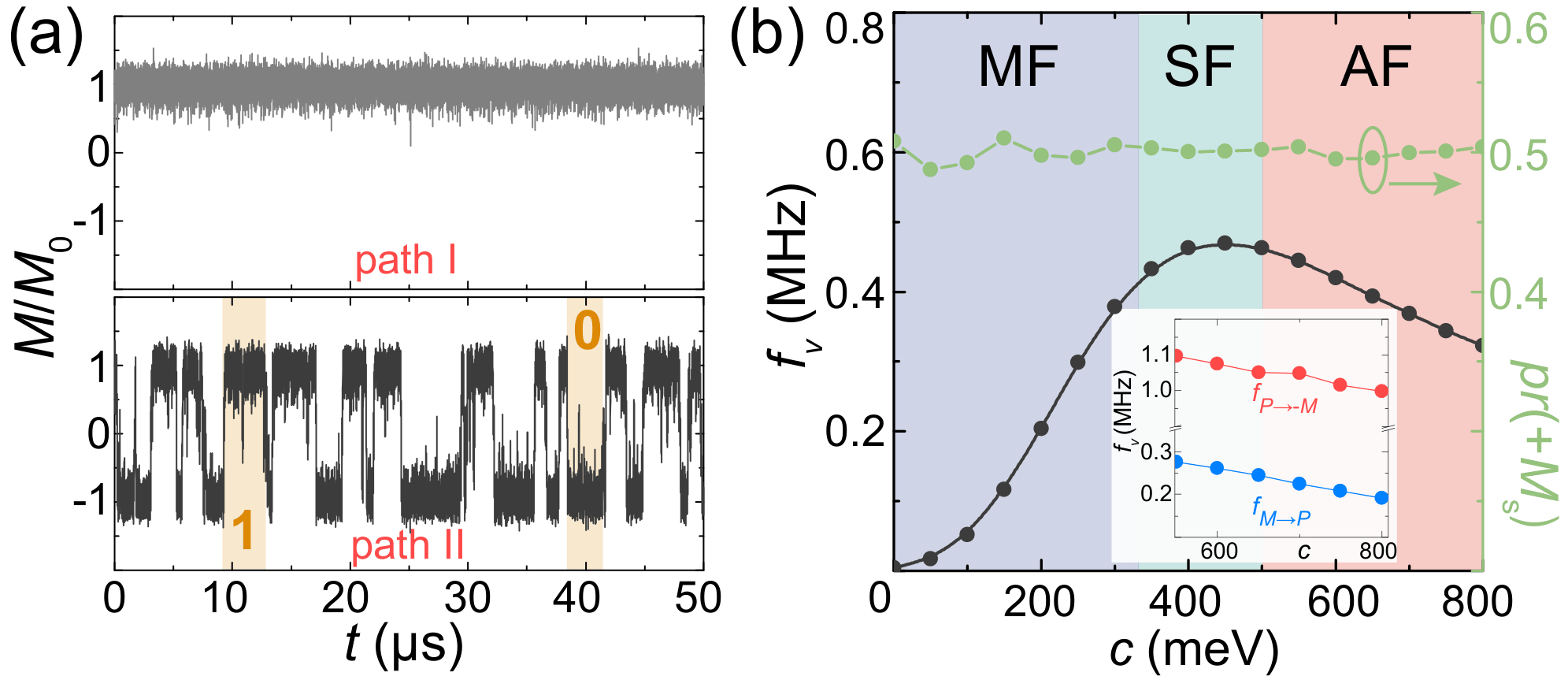}
\caption{Thermally activated magnetic switching in autferroic single-unit-cell limit as the source of TRNG. (a) The simulated telegraph noise of magnetization along the switching path I (obtained with $c=0$ meV) and path II (obtained with $c=550$ meV). (b) The switching rate $f_v$ and time-averaged occupancy probability $pr(+M_s)$ as a function of the ME coupling coefficient $c$. Inset: the subprocess switching rate $f_{M\to P}$ and $f_{P\to -M}$ in the autferroic region.}
\label{Fig2}
\end{figure}

Generally, the TRNG generation rate is primarily influenced by $F_v/k_BT$, where $k_BT$ is the thermal energy, and the $f_v$ increases with rising temperature for a fixed barrier (here the temperature $T$ is set as 300 K if not noted explicitly). Then the stochastic switching between degenerate binary $\pm M_s$ states is simulated by solving the Landau-Khalatnikov (L-K) equation \cite{23, 24}, as explained in SM \cite{22}. The temporal evolution of magnetic order $M(t)$ is recorded at $1$ ns intervals (i.e. the record frequency $f_r=1$ GHz), showing characteristics of telegraph noise [Fig.~\ref{Fig2}(a)]. Here, the fluctuation switching time $\tau$ is defined as the mean duration required to reverse the magnetic direction. As shown in Fig.~\ref{Fig2}(a), the simulated magnetization reveals no flips along the path I within $50$ $\mu$s. The estimated frequency $f_v=\tau^{-1}$ is lower than $100$ Hz according to longer time simulation (not shown here). Instead, the rate is significantly enhanced to $f_v\sim440$ kHz, due to the reduced effective barrier of path II. Furthermore, the switching rate $f_v$ is calculated as a function of ME coupling $c$, as shown in Fig.~\ref{Fig2}(b). It increases rapidly in the multiferroic phase, reaches the maximum in the single-ferroic phase, and slightly decreases in the autferroic region. This behavior was expected, considering the evolution of effective barrier $F_v$ as shown in Fig.~\ref{Fig1}(d).

In the autferroic phase, the entire switching along the path II proceeds via two sequential subprocesses: first from the magnetic state to the ferroelectric state ($+M \to P$) and then back to the magnetic state ($P \to -M$) [Fig.~\ref{Fig1}(b)]. The switching rate of each subprocess is governed by its respective energy barrier, namely $F_{M\to P}$ for $M\to P$ and $F_{P\to -M}$ for $P\to -M$, which are primarily determined as $F_{M\to P}=F_B-F_M$ and $F_{P \to -M}=F_B-F_P$ ($F_B$: the highest free energy point in the path II, i.e. the barrier). As a cascade reaction, the entire switching rate in the autferroic phase can be derived as: $1/f_v=\frac{1}{2} (1/f_{M\to P}+1/f_{P\to -M})$. For autferroic TiGeSe$_3$ monolayer ($c_M>c_P$), $f_{M\to P}$ exceeds $f_{P\to -M}$ resulting in a substantially faster switching rate in the $P\to -M$ process. Thus, the entire $f_v$ is mostly determined by $f_{M\to P}$, and the role of $f_{P\to -M}$ is only to decrease $f_v$ slightly. Consequently, for the single-ferroic phase, $F_{M\to P}=0$ when $c_M>c_P$, giving the fastest $f_v$.

Moreover, our statistical analysis shows that the probability of the time-averaged occupancy of the $+M_s$ state is $\sim50\%$ in both the single-ferroic and autferroic phases [Fig.~\ref{Fig2}(b)]. This balanced distribution indicates that there is no systematic bias of the switching dynamics. Although the switching rate $f_v$ between binary magnetic states in autferroics is slightly lower than that in the single-ferroics, the metastable ferroelectric phase (i.e. $F_{P\to M}>0$) could afford enhanced functionality in TRNG devices, such as quaternary TRNGs, to be discussed later. Besides the ME coupling strength $c$, the faster switching rate $f_v$ requires the small energy difference between magnetic and ferroelectric phases in autferroics ($\Delta F$), see EM for details. 

\begin{figure}
\centering
\includegraphics[width=0.48\textwidth]{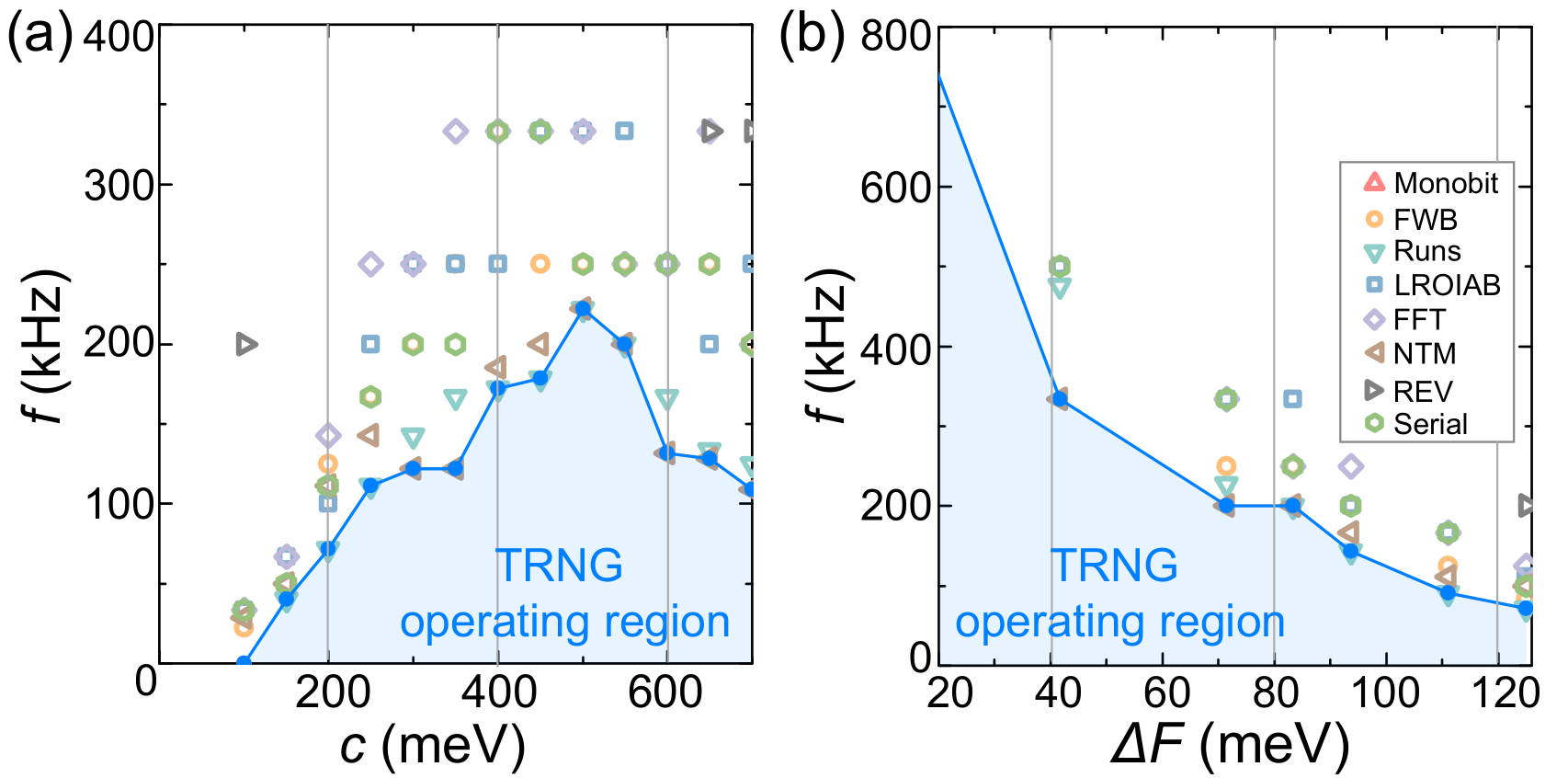}
\caption{The verification of random number bitstream using the NIST Statistical Test Suite. Symbols: the highest frequencies passing different tests. Curves: the minimal of these frequencies is the final working $f_w$, which can pass all eight tests. (a) The testing results as a function of ME coupling $c$. (b) The testing results as a function of free-energy difference $\Delta F$ between the ferroelectric and magnetic phases.}
\label{Fig3}
\end{figure}

The enhanced switching rate $f_v$, the randomness of these switching signals requires further verification via NIST Statistical Test \cite{26}, which includes eight core tests. The working random-number generation rate $f_w$ is determined by the minimum of highest sampling frequencies that successfully pass different tests. When the switching follows the path II in the autferroics, the ME coupling enhances $f_w$ by several orders of magnitude [Fig.~\ref{Fig3}(a)], increasing from under $100$ Hz (at $c=0$ meV) to above 200 kHz maximum [at $c=550$ meV, notably the value where the barrier is lowest, in Fig.~\ref{Fig1}(d)]. As expected, $f_w$ is in the same order of magnitude as but lower than the thermal-fluctuation switching rate $f_v$ ($=440$ kHz) at $c=550$ meV. Furthermore, $f_w$ can be further increased to $\sim 1.1$ MHz when the magnetic and ferroelectric phases become energy-degenerated (i.e. $\Delta F=0$, giving $f_v=2.1$ MHz) [Fig.~\ref{Fig3}(b)]. Therefore, to realize high-performance autferroics-based TRNG, candidate systems should exhibit optimized ME coupling strength and minimal energy disparity between the magnetic and ferroelectric phases.

Having established the principles of bias-free switching, sub-MHz generation rates, and high randomness for ATJ-based TRNGs, we now turn to discuss a realistic physical issue, i.e. domains commonly existing in ferroic materials. In finite-size but single domain device, the switching barrier scales linearly with the free layer volume $V$ ($f_v \propto V$) \cite{safranski2021demonstration}. Therefore, scaling down its volume naturally accelerates the fluctuation rate, but it faces the physical limits such as demagnetization and depolarization field effects. However, this proportionality should vanish in the multi-domain case \cite{vodenicarevic2017low}, where the reversal is governed by the domain nucleation and domain-wall dynamics \cite{miller1960mechanism,fukumoto2006dynamics}. In the framework of Landau theory, the domain-wall effect can be simulated by introducing the spatial gradient terms: $\lambda_M(M_i-M_j)^2$ and $\lambda_P(P_l-P_k)^2$ \cite{18}, with two additional coupling coefficients $\lambda_M$ and $\lambda_P$ between the nearest-neighbor sites $i$ and $j$ ($l$ and $k$). Without loss of generality, for the simulations below both $\lambda_M$ and $\lambda_P$ are set to $2$ meV.

\begin{figure}
\centering
\includegraphics[width=0.45\textwidth]{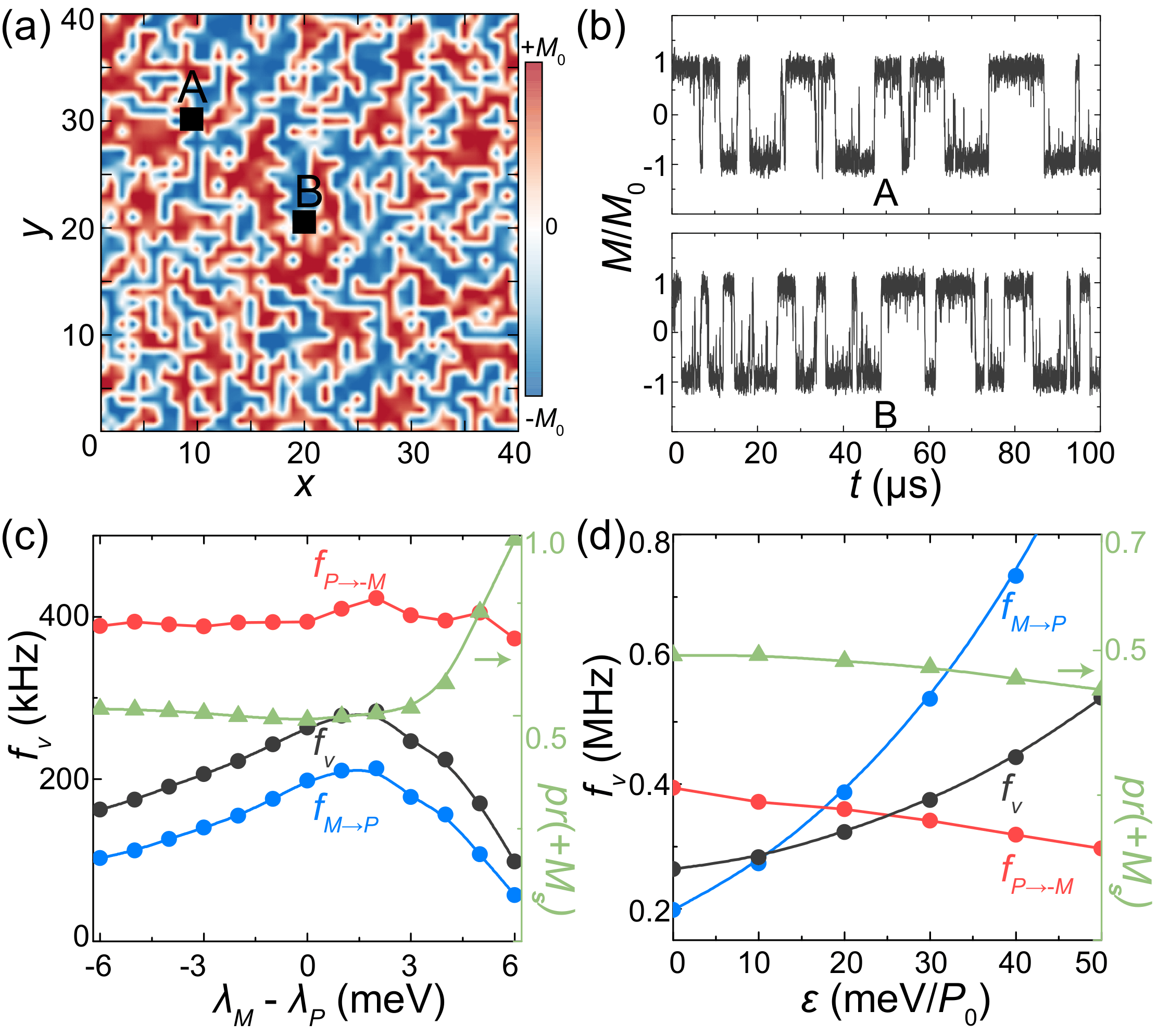}
\caption{Thermally activated magnetic switching in autferroics with multiple domains. (a) A MC snapshot (taken at $t=50$ $\rm{\mu}$s) of magnetic domains. For better resolution ratio, only part of the lattice is shown. (b) Telegraph noise of magnetization in locations A and B as indicated in (a), the footprints of the contacts in Fig.~\ref{Fig1}(c). (c) Thermally activated switching rate $f_v$’s and randomness $pr(+M_s)$ as a function of $\lambda_M-\lambda_P$. (d) Switching rate $f_v$ and randomness $pr(+M_s)$ as a function of $\varepsilon$.}
\label{Fig4}
\end{figure}

With performing a time-dependent Monte Carlo (MC) simulation \cite{27}, the distribution of local magnetization at $t=50$ \rm{$\mu$}s is shown in Fig.~\ref{Fig4}(a), where the white regions represent the domain walls between $+M$ and $-M$. The effect of couplings $\lambda_M$ and $\lambda_P$ on the fluctuation-driven switching rate is further explored by tracking $M(t)$ in locations A and B as shown in Fig.~\ref{Fig4}(a). Both locations exhibit random switching behaviors, with an average switching rate $\sim260$ kHz [Fig.~\ref{Fig4}(b)], which is somewhat lower than that in a single unit cell case. It is reasonable considering the drag effects from neighboring couplings, see EM for details. Therefore, stronger neighboring coupling can lead to higher nucleation barrier, and the lower random switching rate. Other factors of domain wall pinning, such as possibly Peierls-Nabarro-type from the lattice, and further by defects or disorder, in autferroics can also influence the TRNG performance, particularly by slowing the switching rate at high defect concentrations.

In autferroics-based TRNGs, the rate of fluctuation switching and random number generation can be further tuned by applying an external electric field $\varepsilon$. In principle, the free energy of the mediating ferroelectric states (i.e., $F_P$) can be modified by the field ($\pm \varepsilon P_s$ for $\mp P_s$), while the free energy of magnetic phase (i.e. $F_M$) remains unaffected, see EM for details. This field dependent-$f_v$ can be described by the Arrhenius-N\'eel law: $f_x=f_{x_0}{\rm{exp}}(-\varepsilon\cdot P_s/k_BT)$ \cite{28}, where $x=M\to P$ or $P\to -M$ and the subscript $0$ denotes the case without electric field. As a result, when $F_M>F_P$, the total switching frequency $f_v$ increases with electric field strength, as shown in Fig.~\ref{Fig4}(d). Meanwhile, the time-averaged occupancy probability $pr(+M_s)$ is still close to $0.5$ [Fig.~\ref{Fig4}(d)], since the $\pm M_s$ states remain unaffected by the electric field. Hence, the generation rate of autferroics-based TRNG could be enhanced by the electric field, while maintaining its excellent randomness and unaffected magnetoresistance.

\begin{figure}
\centering
\includegraphics[width=0.45\textwidth]{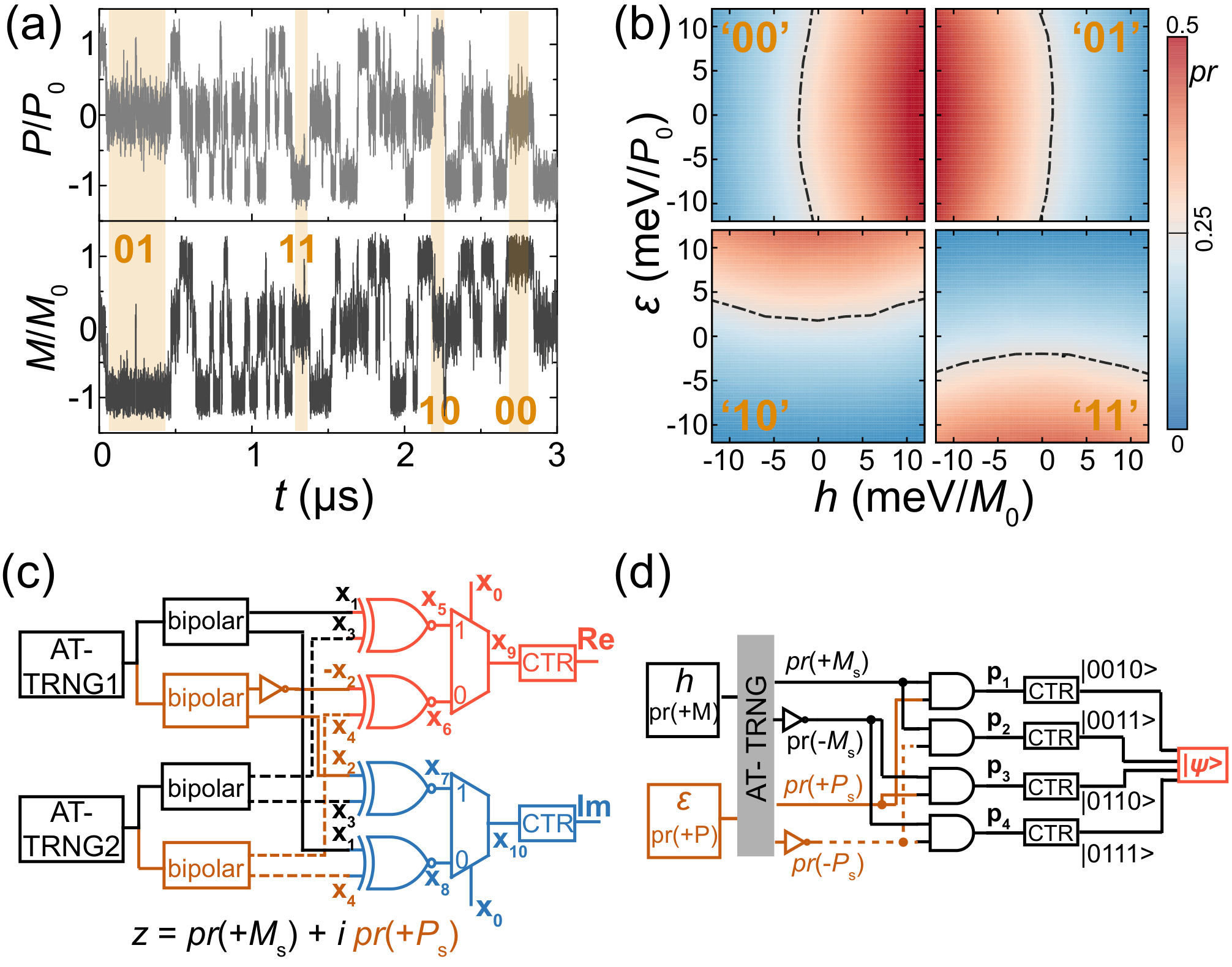}
\caption{Quaternary randomness in near-symmetric autferroics with $c_M\approx c_P$. (a) Calculated telegraph noise for the magnetization and polarization fluctuations. Quaternary states (``$00$'', ``$01$'', ``$10$'', and ``$11$'') are marked as orange. (b) Time-averaged occupancy probabilities of the four quaternary states under combined electric and magnetic field tuning; color map is shown on the right. (c-d) Proposed stochastic circuits for (c) complex-number arithmetic and (d) simulation of probability distributions of quantum mixed states}.
\label{Fig5}
\end{figure}

Above study is based on the case of $c_M>c_P$, in which the magnetic signal is utilized as the source of binary data stream of TRNG and electric field can be the accelerating force. For the counterpart with $c_P>c_M$, the electric signal can be utilized as the binary data stream, while magnetic field is the accelerating force. While for those autferroics with similar values of $c_M$ and $c_P$, the special quaternary data stream is available. In this case, the proximate energies of ferroelectric and magnetic phases lead to similar switching rates between the $M\to P$ and $P\to M$ processes (Fig. S2 \cite{22}). Furthermore, each of the four ferroic states: ($+M_s$, $0$), ($-M_s$, $0$), ($0$, $+P_s$), and ($0$, $-P_s$), is occupied with approximate equal probability of $25\%$, which can be mapped to the quaternary symbols ``$00$'', ``$01$'', ``$10$'', and ``$11$'', respectively [Fig.~\ref{Fig5}(a)].

Consequently, a weak electric (magnetic) field primarily influences the ferroelectric (magnetic) phase in autferroics while having a limited effect on its magnetism (ferroelectricity) [Fig.~\ref{Fig5}(b)]. For instance, applying an electric field along the $+P_s$ direction significantly increases the probability of ($0$, $+P_s$) state and reduces the probability of (0, $-P_s$) state synchronously, whereas the probabilities of the ($\pm M_s$, $0$) states remain largely unaffected. This selective coupling allows independent tuning of the state occupation probabilities, thereby realizing four-element probabilistic bits ($p$-bits), which constitutes the key building blocks of $p$-computers \cite{1}. Although in those type-I multiferroics the quaternary states ($\pm M_s$, $\pm P_s$) are also available, their response to external fields is double-degenerated, i.e. the electric field can only increase the probabilities of the ($\pm M_s$, $\pm P_s$) states and reduce the other two.

Given its near-independent control of ($\pm M_s$, $0$) and ($0$, $\pm P_s$) states, an autferroic ATJ containing two $p$-bits can effectively encode a complex number, wherein the two ferroic orders can be utilized to represent the real and imaginary parts via independent binary encoding. For example, the sampled magnetic binary states in autferroic TRNG represent the real part, with ($+ M_s$, $0$) and ($-M_s$, $0$) defined as ‘$1$’ and ‘$0$’, which can be distinguished through ATJ’s TMR. Similarly, the ferroelectric state provides the imaginary part. Hence, the values of real and imaginary part can be individually binary encoded by $pr[(+M_s, 0)]$ and $pr[(+P_s, 0)]$, which are both close to $0.5$ without external fields. Moreover, complex-number scaled multiplication can be physically realized using a probabilistic circuit ($p$-circuit) consisting of only two ATJ units [Fig.~\ref{Fig5}(c)].

The $pr[(+M_s, 0)]$ and $pr[(+P_s, 0)]$ in each unit are first converted into signed values via bipolar coding $x=2pr-1$. After the bipolar coding, the complex scaled multiplication $z_1\cdot  z_2 = (x_1 + i x_2)(x_3 + i x_4)$ with all four $p$-bits ($x_1$, $x_2$, $x_3$, $x_4$) is performed using just four stochastic multipliers and two scaled adders/subtractors according to Re($z_1\cdot  z_2$) $= x_1 x_3+ (-x_2 x_4) = x_5 + x_6$, and Im($z_1\cdot  z_2$) $= x_1 x_4 + x_2 x_3 = x_7 + x_8$. Specifically, the four stochastic multiplications, $x_1x_3 = x_5$, $x_2x_4 = x_6$, $x_1x_4 = x_7$, and $x_2x_3 = x_8$, are implemented using exclusive NOR (XNOR) gates, while the required negation for the term $-x_2x_4$ is performed by a single NOT gate on one of the multiplicand streams prior to its XNOR operation [Fig.~\ref{Fig5}(c)]. Scaled addition and subtraction are then carried out using 2-to-1 multiplexers selected by an independent TRNG stream $x_0$ [Fig.~\ref{Fig5}(c)], producing $x_9=(x_1x_3+x_2x_4)/2$ and $x_{10}=(x_1x_4+x_2x_3)/2$ for the real and imaginary parts of the product, respectively. The $1/2$ scaling factor intrinsic to multiplexer (MUX) based stochastic adders is easily corrected digitally via counter-based normalization \cite{29}. Consequently, autferroic-based $p$-bit units can substantially simplify the overall $p$-circuit architecture compared with conventional designs relying on separate TRNGs.

Beyond the classical probabilistic computing, the same physical encoding of tunable quaternary probabilities in autferroic-based TRNGs can be used to simulate the probability distribution of a quantum mixed state, which are enabled by its strong seesaw-type ME coupling [Fig.~\ref{Fig5}(d)]. Specifically, the quantum wave function realized by a single antiferroic unit is expressed as $\left | \psi  \right \rangle =\left (\sqrt{pr(+M_s)} \left | 00  \right \rangle + \sqrt{pr(-M_s)} \left | 01  \right \rangle \right )\otimes \left (\sqrt{pr(+P_s)} \left | 10  \right \rangle + \sqrt{pr(-P_s)} \left | 11  \right \rangle \right )$. These four states are measured subsequently mapped to a quaternary encoding scheme. Different from the aforementioned binary encoding scheme [Fig.~\ref{Fig5}(c)], now the probabilities of each states are approximate $0.25$ without field. Expanding the state yields $\left | \psi  \right \rangle =\sqrt{p_1} \left | 0001  \right \rangle + \sqrt{p_2} \left | 0011  \right \rangle + \sqrt{p_3} \left | 0111  \right \rangle + \sqrt{p_4} \left | 0111  \right \rangle$ with joint probabilities $p_1 = pr(+M_s)pr(+P_s)$, $p_2 = pr(+M_s)pr(-P_s)$, $p_3 = pr(-M_s)pr(+P_s)$, and $p_4 = pr(-M_s)pr(-P_s)$ naturally computed by stochastic AND gates in the p-circuit [Fig.~\ref{Fig5}(d)]. Because the original physical states are pairwise orthogonal due to the seesaw-type ME coupling, the resulting two-qubit basis states remain orthogonal in the Hilbert space ($\left \langle jk  | kl  \right \rangle =\delta_{jk}\delta_{kl}$), thereby satisfying the formal requirements of quantum state representation \cite{30}. Consequently, a single autferroic-based TRNGs unit may provide a compact, field-tunable source for simulating probability distributions associated with quantum mixed states.

In summary, strong seesaw-type ME coupling in autferroics results in a lower switching barrier between the ground state ferroic order. This enables rapid thermal fluctuations of spontaneous magnetization (or polarization) without the sacrifice of its swing magnitude, making autferroic-based tunneling junction superior to conventional magnetic tunneling junction for high-performance TRNG devices. In the asymmetric case, e.g. magnetic-preferred $c_M>c_P$, thermally activated fluctuations between degenerate $±M_s$ states are predicted to exhibit fast switching rate and high-quality binary randomness. Moderate external fields can boost the TRNG bitrate. In the symmetric case, i.e. $c_M \approx c_P$, autferroics can be naturally implemented into a field-tunable quaternary TRNG, which can facilitate compact stochastic circuits for complex-number arithmetic and simulation of probability distributions of quantum mixed states. In short, the distinct physics of autferroics provides a potentially transformative platform for next-generation computing paradigms.

\bibliography{ref}
\newpage
\onecolumngrid
\section{End Matter}
\twocolumngrid
{\it EM1. Various physical quantities used throughout the work.}
\begin{table}[H]
\centering
\resizebox{0.39\textwidth}{!}{
\begin{tabular}{llcc}
\hline
\hline
physical quantity &$~$    &meaning \\
\hline
\multirow{3}*{magnetic} &$M$  &magnetic order parameter\\
\cline{2-3}
$~$ &$M_0$   &\makecell[c]{spontaneous magnetic moment\\ of pure magnetic system}  \\
\cline{2-3}
$~$ &$M_s$   &saturation magnetization  \\
\hline
\multirow{3}*{ferroelectric} &$P$   &ferroelectric order parameter  \\
\cline{2-3}
$~$ &$P_0$   &\makecell[c]{spontaneous polarization of pure \\ferroelectric system}  \\
\cline{2-3}
$~$ &$P_s$   &saturation polarization  \\
\hline
\multirow{3}*{coupling} &$c$   &magnetoelectric coupling strength  \\
\cline{2-3}
$~$ &$c_M$   &\makecell[c]{critical threshold for \\magnetism: $c_M\equiv2F_M/P_0^2M_0^2$} \\
\cline{2-3}
$~$ &$c_P$   &\makecell[c]{critical threshold for \\ferroelectricity: $c_P\equiv2F_P/P_0^2M_0^2$}  \\
\hline
\multirow{7}*{energy} &$F$   &Landau free energy at $T=0$ K  \\
\cline{2-3}
$~$ &$F_M$   &pure magnetic switching barrier  \\
\cline{2-3}
$~$ &$F_P$   &pure ferroelectric switching barrier  \\
\cline{2-3}
$~$ &$F_v$   &\makecell[c]{minimum switching barrier between\\ ($+M_s$, $0$) and ($-M_s$, $0$)}  \\
\cline{2-3}
$~$ &$F_B$   &\makecell[c]{highest free energy point between\\ magnetic and ferroelectric phases}  \\
\cline{2-3}
$~$ &$F_{M\to P}$   &\makecell[c]{switching barrier from magnetic to\\ ferroelectric phases}  \\
\cline{2-3}
$~$ &$F_{P\to M}$   &\makecell[c]{switching barrier from ferroelectric\\ to magnetic phases}  \\
\hline
\multirow{5}*{frequency} &$f_v$   &switching rate: $f_v=\tau^{-1}$  \\
\cline{2-3}
$~$ &$f_{M\to P}$   &\makecell[c]{subprocess switching rate from\\ magnetic to ferroelectric phases}  \\
\cline{2-3}
$~$ &$f_{P\to M}$   &\makecell[c]{subprocess switching rate from\\ ferroelectric to magnetic phases}  \\
\cline{2-3}
$~$ &$f_r$   &record frequency during simulation  \\
\cline{2-3}
$~$ &$f_w$   &\makecell[c]{working random-number\\ generation rate}  \\
\hline
\multirow{2}*{gradient} &$\lambda_M$   &\makecell[c]{magnetic spatial gradient\\ between nearest-neighbors}  \\
\cline{2-3}
$~$ &$\lambda_P$   &\makecell[c]{ferroelectric spatial gradient\\ between nearest-neighbors}  \\
\hline
\multirow{6}*{miscellaneous} &$T$   &temperature  \\
\cline{2-3}
$~$ &$pr$   &probability  \\
\cline{2-3}
$~$ &$\tau$   &fluctuation switching time  \\
\cline{2-3}
$~$ &$V$   &volume  \\
\cline{2-3}
$~$ &$\varepsilon$   &electric field  \\
\cline{2-3}
$~$ &$h$   &magnetic field  \\

\hline
\hline
\end{tabular}
}
\label{Table1}    
\end{table}

{\it EM2. More details of phase diagram.}
In the ground state phase diagram, the regions of multiferroic and autferroic phases are determined solely by the strength of ME coupling $c$. In the weak ME coupling side (i.e. $c<c_M$ and $c<c_P$), the ground state is the type-I multiferroic, allowing the coexistence of ferroelectricity and magnetism. In the strong ME coupling side (i.e. $c>c_M$ and $c>c_P$), the autferroic phase appears as the ground state, and four wells (ferroelectric $vs.$ magnetic) exist in its energy landscape, as shown in Fig.~\ref{Fig1}(a). In the middle region (between $c_M$ and $c_P$), an unexpected single-ferroic phase emerges as the bridge between the multiferroic and autferroic ones [Fig.~\ref{Fig1}(d)], where one ferroic order dominates the global minimum of energy landscape while another order occupies the saddle point.

\begin{figure}
\centering
\includegraphics[width=0.45\textwidth]{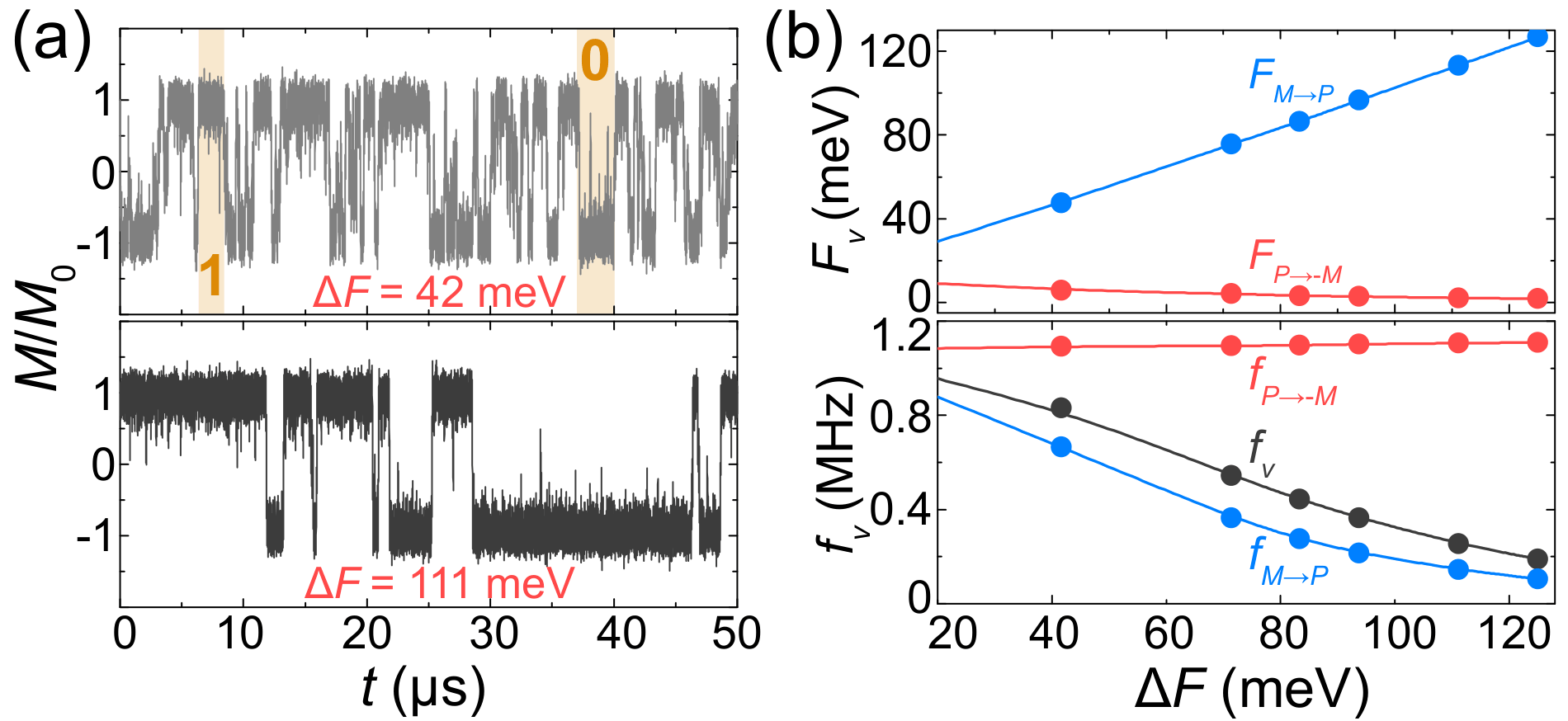}
\caption{Thermally activated magnetic switching in autferroics, as the source of TRNG. (a) The telegraph noise of magnetization at two cases of different $\Delta F=F_M-F_P$, representing free-energy difference between the ferroelectric and magnetic phases, while $c$ is fixed to $550$ meV. (b) The switching barrier $F_v$’s and switching rate $f$’s as function of $\Delta F$.}
\label{Fig2-em}
\end{figure}

{\it EM3. More details of effects of $\Delta F$.}
The switching barrier $F_v$ along path II is also influenced by the energy difference between magnetic and ferroelectric phases in autferroics, i.e. $\Delta F=F_M-F_P=(c_M-c_P)/2$ when $c_M>c_P$ [or $\Delta F=F_P-F_M=(c_P-c_M)/2$ when $c_P>c_M$]. Thus, $\Delta F$ can further affect the switching rate $f_v$. To investigate effect of $\Delta F$ on switching rate, we calculate the time evolution of $M(t)$ for different $\Delta F$’s by adjusting the energy potential depth of magnetic state while keeping all other parameters unchanged. Our results show that magnetic switching rate decreases rapidly as $\Delta F$ increases [Fig.~\ref{Fig2-em}(a)]. Specifically, the switching rate $f_{M\to P}$ remains largely unaffected, while $f_{P\to -M}$ as the main contributor to the entire $f_v$ decreases significantly due to the reduced barrier $F_{P\to M}$ [Fig.~\ref{Fig2-em}(b)]. 

\begin{figure}
\centering
\includegraphics[width=0.28\textwidth]{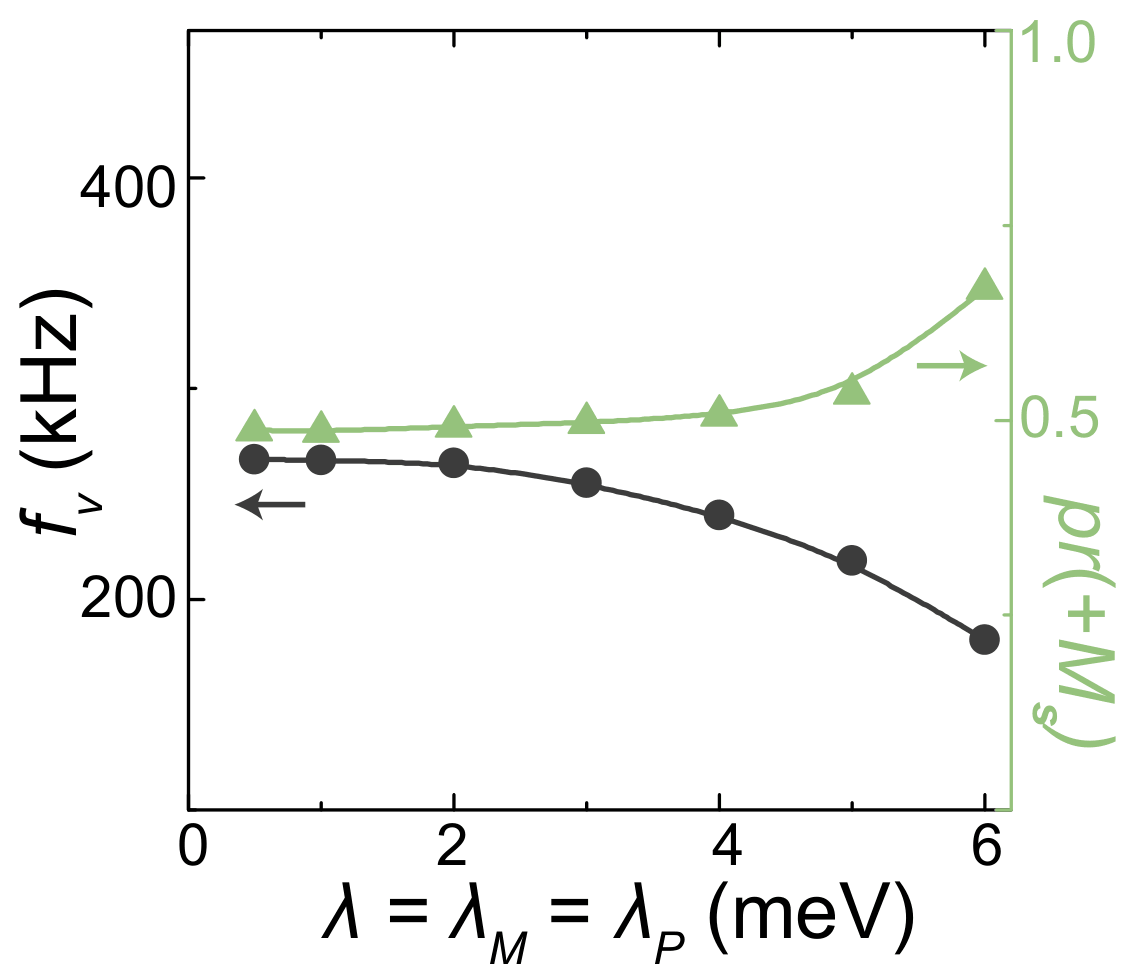}
\caption{Thermally activated switching rate $f_v$’s and randomness $pr$(+$M_s$) as a function of $\lambda$.}
\label{fig6-em}
\end{figure}

{\it EM4. More details on the effect of neighboring couplings.}
In the multi-domain system, the reversal is governed by local domain nucleation barrier $F_n$ and domain-wall motion velocity $v$ \cite{miller1960mechanism,fukumoto2006dynamics}. Also these two factors are both related to neighboring couplings ($F_n \propto \lambda$, $v \propto e^{-\sqrt{\lambda}}$). Here, we first considered the case where $\lambda = \lambda_M = \lambda_P$ while keeping all other parameters unchanged. The calculated $f_v$ as a function of $\lambda$ at room temperature is shown in Fig.~\ref{fig6-em}. As expected, the $f_v$ decreases with increasing $\lambda$ due to the higher nucleation barrier resulting from increased domain-wall energy \cite{miller1960mechanism,fukumoto2006dynamics}. 

In autferroics, the magnetic and ferroelectric orders originate from different sublattices, and their domain walls are mutually exclusive due to the strong seesaw-type ME coupling. Therefore, the effect of differing $\lambda_M$ and $\lambda_P$ values on the switching rate is also investigated [Fig. ~\ref{Fig4}(c)]. In the $c_M>c_P$ case, the entire switching rate $f_v$ along path II is primarily influenced by the slower $M\to P$ process. Thus, the switching rate of the $M\to P$ process declines rapidly when $\lambda_M$ is raised, as the stability of the magnetic long-range order is enhanced. Conversely, the rate of the $P\to M$ process initially decreases before increasing at higher $\lambda_M$. As a result, the total switching rate reaches its maximum at around $\lambda_M=4$ meV ($\lambda_M–\lambda_P=2$ meV) before starting to decline, as shown in Fig.~\ref{Fig4}(c). Furthermore, our statistical analysis reveals that the probability of the time-averaged occupancy of the $+M_s$ state significantly deviates from equilibrium [for example $pr(+M_s)>0.6$] within the measuring duration ($100$ $\rm{\mu}s$) when $\lambda_M>4$ meV [Fig.~\ref{Fig4}(c)]. These results suggest that strong magnetic exchange coupling in the $c_M>c_P$ case suppresses fluctuation-driven switching rates, partially froze the system in the $+M_s$ state, and thus reduces its overall randomness. Additionally, while the value of $\lambda_P$ slightly weakens the switching rate of the $M\to P$ process, its effects on the $P\to M$ process are minimal. 

Hence, the multi-domain system may also approach or achieve the high-frequency MHz regime, when the neighboring coupling is optimized to lower the nucleation barrier. In this case of large devices, effects of demagnetization or depolarization field are rather limited.

\begin{figure}
\centering
\includegraphics[width=0.43\textwidth]{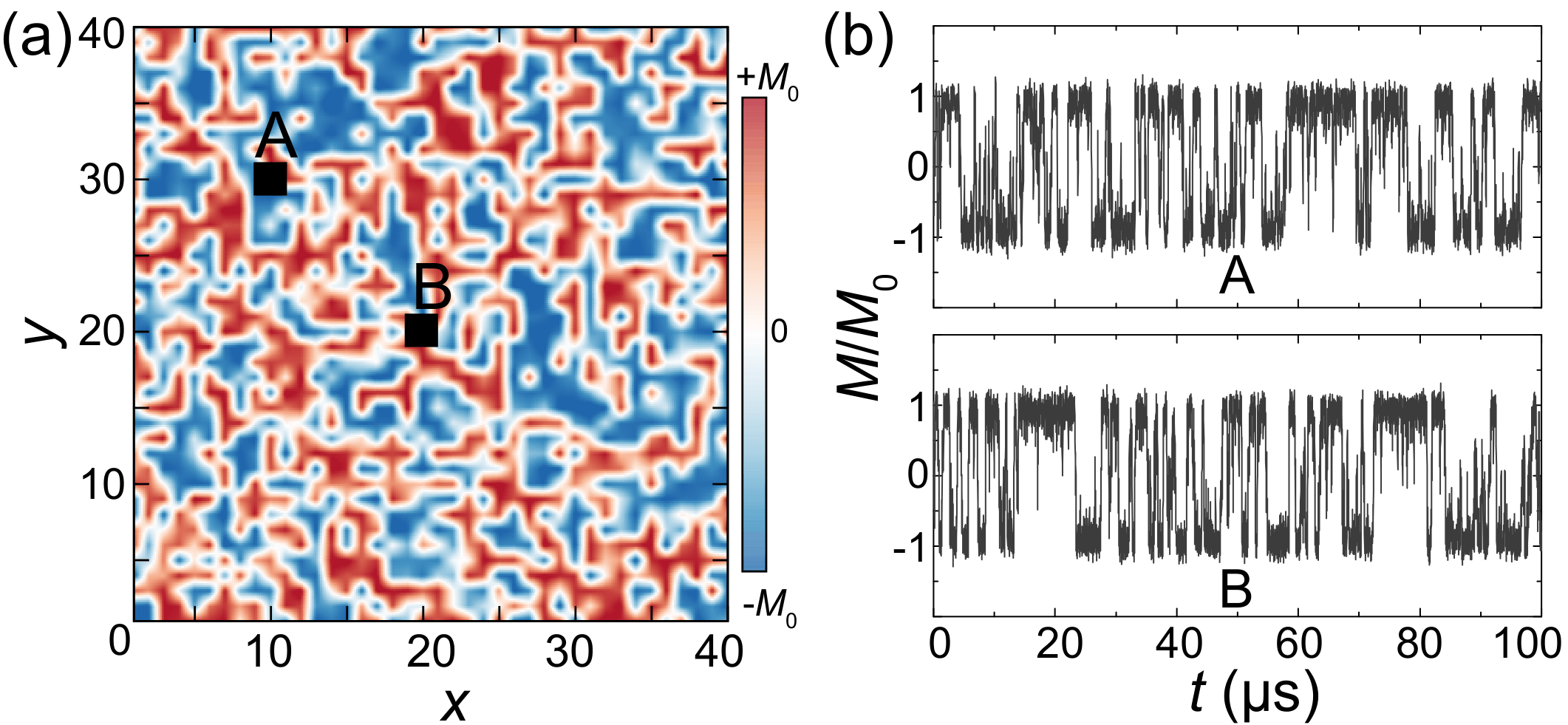}
\caption{Thermally activated magnetic switching in autferroics with external field $\varepsilon$. (a) A MC snapshot (taken at $t=50$ $\rm{\mu}$s) of magnetic domains. For better resolution ratio, only part of the lattice is shown. (b) Telegraph noise of magnetization in locations A and B as indicated in (a).}
\label{fig4-em}
\end{figure}

{\it EM5. More details of effects of external fields.}
We re-perform time-dependent MC simulations with external electric fields. As shown in Fig.~\ref{fig4-em}(a), the magnetic domain walls (i.e. the mediating ferroelectric state) become wider under electric fields. Additionally, the fluctuation-driven magnetic switchings are accelerated [Fig.~\ref{fig4-em}(b)]. Specifically, the energy barrier for the $M\to P$ process is substantially lowered by the electric field, although the switching barrier in the $P\to -M$ process increases somewhat.

\end{document}